\documentclass{elsart}
\usepackage{bm}
\usepackage{graphicx}

\def\d{{\rm d}}
\def\dk{{\rm d}k}
\def\dt{{\rm d}t}

\begin{document}

\journal{Physica D}
\pubyear{2004}
\volume{}

\begin{frontmatter}

\title{Nonlinear transfer and spectral 
distribution of energy in $\alpha$ turbulence}

\author{Chuong V. Tran}

\address{Department of Mathematical and Statistical Sciences, University 
of Alberta, Edmonton, Alberta, Canada, T6G 2G1}

\ead{chuong@math.ualberta.ca}

\begin{abstract} 
Two-dimensional turbulence governed by the so-called $\alpha$ turbulence 
equations, which include the surface quasi-geostrophic equation 
($\alpha=1$), the Navier--Stokes system ($\alpha=2$), and the governing 
equation for a shallow flow on a rotating domain driven by a uniform
internal heating ($\alpha=3$), is studied here in both the unbounded
and doubly periodic domains. This family of equations conserves two
inviscid invariants (energy and enstrophy in the Navier--Stokes case),
the dynamics of which are believed to undergo a dual cascade. It is
shown that an inverse cascade can exist in the absence of a direct 
cascade and that the latter is possible only when the inverse transfer
rate of the inverse-cascading quantity approaches its own injection
rate. Constraints on the spectral exponents in the wavenumber ranges
lower and higher than the injection range are derived. For
Navier--Stokes turbulence with moderate Reynolds numbers, the
realization of an inverse energy cascade in the complete absence of a
direct enstrophy cascade is confirmed by numerical simulations. 
\end{abstract}


\end{frontmatter}

\section{Introduction}

Incompressible fluid turbulence in two dimensions has always been a
subject of interest because of its relative simplicity and broad 
applications in geophysical problems. Although the theory of 
two-dimensional (2D) turbulence is not strictly applicable to a 
geophysical fluid, the third (vertical) dimension in many geophysical 
systems can nevertheless be ``removed'', and the resulting 2D equations
are thought to be quantitatively adequate for describing the 
large-scale horizontal motion.
  
Some systems in this category, which receive considerable attention, 
are the surface quasi-geostrophic (SQG) equation, describing the motion 
of a rotating stratified fluid \cite{Blumen78,Held95}; the ``rotating 
shallow flow'' (RSF) equation, governing the motion of a thin fluid on 
a rotating domain \cite{Weinstein89}; and the Charney--Hasegawa--Mima 
equation, governing the potential vorticity of an equivalent-barotropic 
fluid \cite{Charney48,Hasegawa78,Hasegawa79}. Except possibly for the last 
equation, these systems, together with the more familiar 2D 
Navier--Stokes (NS) equations, can be studied within a unified framework 
by a family of models (known as the $\alpha$ turbulence equations) indexed 
by a parameter $\alpha\in(0,\infty)$ \cite{Pierrehumbert94}. Like the 
2D NS system, the nonlinear transfer of $\alpha$ turbulence conserves two 
quadratic invariants: $\Psi_\alpha=\int_0^\infty\Psi_\alpha(k)\,\dk$ 
and $\Psi_{2\alpha}=\int_0^\infty\Psi_{2\alpha}(k)\,\dk$, where 
$\Psi_{\theta}(k)= k^{\theta}\Psi(k)$, $\Psi(k)$ is the streamfunction 
power spectrum, $\theta$ is a real number, and $k$ is the wavenumber.
Note that $\Psi_2(k)$ is the kinetic energy spectrum, and $\Psi_2$ is
the kinetic energy. 

The statistical dynamics of unbounded 2D NS turbulence are believed to 
achieve a quasi-steady state which simultaneously exhibits a direct 
enstrophy cascade to high wavenumbers and an inverse energy cascade to 
low wavenumbers. According to the classical theory formulated by 
Kraichnan \cite{Kraichnan67,Kraichnan71}, Leith \cite{Leith68}, and 
Batchelor \cite{Batchelor69} (hereafter referred to as KLB), if the 
system is excited around a (forcing) wavenumber $s$, the dual cascade 
is supposed to carry virtually all of the energy input to ever-lower 
wavenumbers, down to wavenumber $k=0$, and the enstrophy input to a 
high wavenumber $k_\nu\gg s$, around which the enstrophy is dissipated. 
Analyses based on Kolmogorov's phenomenology predict that the energy 
spectrum scales respectively as $k^{-5/3}$ and $k^{-3}$ in the energy- 
and enstrophy-cascading ranges (also known as inertial ranges). The KLB 
theory, when applied to $\alpha$ turbulence, implies the possibility
that $\Psi_\alpha$ cascades to low wavenumbers and that $\Psi_{2\alpha}$ 
cascades to high wavenumbers. These (possible cascading) invariants 
correspond to different physical quantities for different flows. For 
SQG turbulence, the direct-cascading candidate is the energy. For RSF 
turbulence, the direct-cascading candidate is the palinstrophy. When 
$\alpha=4$, the enstrophy becomes the inverse-cascading candidate. NS 
turbulence features the energy and enstrophy as the possible inverse- 
and direct-cascading quantities, respectively.

Despite numerical evidence confirming the realization of an inverse 
energy cascade and of the Kolmogorov--Kraichnan $k^{-5/3}$ spectrum
\cite{Boffetta00,Borue94,Frisch84,Maltrud91,Tran03c}, the dual 
cascade remains conjectural. No convincing evidence exists of a direct 
enstrophy cascade. Rather, there exist some negative 
results with respect to this problem. In particular, for a bounded 
system in equilibrium, a direct enstrophy cascade is excluded 
\cite{Tran02a}. For an unbounded system, a direct enstrophy cascade 
is prohibited for all weak inverse energy cascades \cite{Tran03c}. 
By ``weak'', it is understood that the inverse cascades do not carry 
virtually all of the energy input to the large scales. This behaviour
is presumably the dynamical behaviour for finite-Reynolds number 
turbulence. The questions then arise whether the inverse energy 
cascade can become ``strong'' in the limit of infinite Reynolds 
number and whether a direct enstrophy cascade can then be realizable.

This study generalizes various previous results for the 2D NS system 
\cite{Tran03a,Tran03c,Tran02a} to $\alpha$ turbulence and re-examines 
the dual-cascade hypothesis under the general framework of the latter 
case. Also, some new results on related issues, such as the 
non-conservative transfer of quadratic quantities other than the 
invariants, are reported. First, the derivations of some a priori 
estimates for the dissipation terms of the quadratic invariants are 
presented. Second, the result 
of Tran and Shepherd \cite{Tran02a} is generalized to the present 
problem, showing that under the influence of a viscous operator, the 
invariants satisfy $\Psi_{2\alpha}\le s^{\alpha}\Psi_\alpha$. Third, 
the conservative transfer of the quadratic invariants and the 
non-conservative transfer of other quadratic quantities are examined. 
The former consideration is essentially a reformulation of the 
dual-cascade hypothesis. It is shown that even in the limit of infinite
Reynolds number, the inverse cascade of $\Psi_\alpha$ could become 
strong, carrying virtually all of the injection of $\Psi_\alpha$ to 
ever-larger scales, in the complete absence of a direct cascade 
of $\Psi_{2\alpha}$. This study does not attempt an alternative 
to the classical dual-cascade theory, but is concerned with the
plausibility of the dual-cascade hypothesis. A constraint on the 
spectral slopes for fluids in equilibrium is obtained by extending 
the derivations of Tran and Bowman \cite{Tran03a,Tran03c} to the 
present case. Finally, the results from numerical simulations of
NS turbulence are presented, confirming the realization of an inverse 
energy cascade in the absence of a direct enstrophy cascade. 
These numerical results suggest that for high Reynolds numbers, 
the turbulence would become non-direct-cascading by evolving 
through a transient direct-cascading stage. 

The remainder of this paper is organized as follows. Section 2 describes
the equation of motion governing $\alpha$ turbulence and its conservation 
laws. Section 3 presents the derivations of the upper bounds on the 
asymptotic averages of the dissipation of quadratic invariants and the
dynamical constraint $\Psi_{2\alpha}\le s^{\alpha}\Psi_\alpha$. Section 4
reports an analysis concerning the plausibility of a direct cascade in
unbounded fluids. Section 5 features a proof of the creation and depletion 
of quadratic quantities other than the invariants. Section 6 presents a 
constraint on the spectral slopes of the kinetic energy spectrum for 
fluids in equilibrium. Section 7 reports the results from numerical
simulations of 2D NS turbulence, confirming the realization of a weak
inverse energy cascade in the complete absence of a direct enstrophy 
cascade. The paper ends with some concluding remarks in the final section.

\section{Governing equation}

The term ``$\alpha$ turbulence'' refers to a general class of 2D 
incompressible fluid turbulence, in which the conserved scalar 
$(-\Delta)^{\alpha/2}\psi$ is advected in a velocity field with stream 
function $\psi$. Here, $\alpha$ is a positive number, and $\Delta$ is 
the 2D Laplacian. The equation governing the fluid motion is 
\begin{eqnarray}
\label{governing}
\partial_t(-\Delta)^{\alpha/2}\psi+J(\psi,(-\Delta)^{\alpha/2}\psi)
&=&-\nu_\mu(-\Delta)^\mu(-\Delta)^{\alpha/2}\psi+f,
\end{eqnarray}
where $J(\varphi,\phi)=\partial_x\varphi\,\partial_y\phi
-\partial_x\phi\,\partial_y\varphi$, $\nu_\mu>0$ is the generalized 
viscosity coefficient, $\mu\ge0$ is the degree of viscosity, and $f$ 
is the forcing. The stream function $\psi(x,y,t)$ is related to the 
fluid velocity $\bm u$ by $\bm u=(-\partial_y\psi,\partial_x\psi)$. 
This family includes the three physically realizable fluid systems, 
which are the focus of this work. The case $\alpha=1$ corresponds to 
the SQG equation from the geophysical context, describing the motion 
of a stratified fluid with small Rossby and Ekman numbers 
\cite{Blumen78,Held95,Pierrehumbert94}. The familiar 2D NS system
corresponds to $\alpha=2$. The case $\alpha=3$ corresponds to the 
RSF system, governing a shallow flow on a rotating domain with 
uniform internal heating \cite{Weinstein89}. 

The nonlinear term in (\ref{governing}) obeys the conservation laws
\begin{eqnarray}
\label{conservation}
\langle\psi J(\psi,(-\Delta)^{\alpha/2}\psi)\rangle=
\langle(-\Delta)^{\alpha/2}\psi J(\psi,(-\Delta)^{\alpha/2}\psi)\rangle=0,
\end{eqnarray}
where $\langle\cdot\rangle$ denotes the spatial average.
As a consequence, the two quadratic quantities $\Psi_{\alpha}=
\frac{1}{2}\langle|(-\Delta)^{\alpha/4}\psi|^2\rangle$ and 
$\Psi_{2\alpha}=\frac{1}{2}\langle|(-\Delta)^{\alpha/2}\psi|^2\rangle$ 
are conserved by nonlinear transfer.

In this study, $f$ is assumed to be spectrally confined to a finite 
band of the wavenumber $K=[k_1,k_2]$, where $k_1>0$. More restrictive 
conditions on $f$ and specific values of $\mu$ for the particular 
cases to be considered will be stated in due course. In particular, 
$\mu=1$ represents molecular viscosity as in the NS case. The SQG 
system has $\mu=1/2$ as the physical dissipation mechanism; other 
values of $\mu$ have also been considered for numerical purposes in 
the literature \cite{Ohkitani97,Schorghofer00a,Schorghofer00b,Smith02}.
It is not well understood how the dynamics would depend on the degree
of viscosity $\mu$. Nevertheless, there appears to be a belief that 
$\nu_\mu(-\Delta)^\mu$ (for large positive values of $\mu$) globally 
acts on the small scales, hence the name ``small-scale dissipation''.
This belief is associated with the concepts of direct cascade and of 
dissipation range and is incorrect if a direct cascade is not 
realizable. For example, for 2D NS turbulence in equilibrium, with 
$\nu_\mu(-\Delta)^\mu$ ($\mu>0$) the only form of dissipation, the 
dissipation of energy (enstrophy) occurs mainly in the large scales 
(around the forcing scale) \cite{Tran02a}. This example means that 
there exists no dissipation range well separated from the forcing 
scale and that the viscous operator $\nu_\mu(-\Delta)^\mu$ would 
have to be correctly termed ``large-scale dissipation''. In this 
case, the degree of viscosity $\mu$ plays an important role in the 
spectral distribution of energy \cite{Tran03a,Tran02a}.

The problem of nonlinear transfer and spectral distribution of 
quadratic quantities such as energy and enstrophy is best handled in 
wavenumber space. Hence, it is convenient to formulate the present 
problem accordingly. The power spectrum $\Psi_\theta(k)$, which 
represents the power density of $(-\Delta)^{\theta/4}\psi$ (for
real $\theta$) associated with the wavenumber $k$, is defined by
\begin{eqnarray}
\label{Psi(k)}
\Psi_\theta(k)&=&\frac{1}{2}\int_{|\bm\ell|=k}
|(-\Delta)^{\theta/4}\widehat\psi(\bm{\ell})|^2\,\d\bm{\ell}
=\frac{1}{2}\,k^\theta\int_{|\bm\ell|=k}
|\widehat\psi(\bm{\ell})|^2\,\d\bm{\ell},
\end{eqnarray}
where $\widehat\psi(\bm{\ell})$ is the Fourier transform of $\psi$, and 
the integral is over all wavevectors $\bm{\ell}$ having magnitude $k$.
In this paper, most results apply to both unbounded and periodic (bounded)
cases; the integral over continuous wavenumbers is then understood as 
the corresponding sum over discrete wavenumbers for the bounded case. 
The evolution of $\Psi_\alpha(k)$ is governed by 
\begin{eqnarray}
\label{Psi(k)evolution}
\frac{\d}{\dt}\Psi_\alpha(k)&=&T_{\alpha}(k)
-2\nu_\mu k^{2\mu}\Psi_\alpha(k)+F(k).
\end{eqnarray}
Here, $T_{\alpha}(k)$ and $F(k)$ are, respectively, the spectral 
transfer and input functions. (See \cite{Frisch95,Kraichnan67} for 
the familiar NS case.) By virtue of the conservation laws
(\ref{conservation}), the transfer function $T_{\alpha}(k)$ satisfies
\begin{eqnarray}
\label{conservation1}
\int_0^\infty T_{\alpha}(k)\,\dk&=&\int_0^\infty k^\alpha T_{\alpha}(k)\,\dk=0.
\end{eqnarray}

\section{Asymptotic analysis}

In this section, several a priori estimates of some relevant
dynamical quantities are derived, and their implications for power-law
scaling of the energy spectrum are discussed. The results 
obtained include upper bounds on the asymptotic averages of the 
dissipation of $\Psi_\alpha$ and $\Psi_{2\alpha}$ and an upper bound 
on $\Psi_{2\alpha}$ in terms of $\Psi_\alpha$ and the forcing wavenumber.

On multiplying (\ref{governing}) by $\psi$ and 
$(-\Delta)^{\alpha/2}\psi$ and taking the spatial averages 
of the resulting equations, noting from the conservation laws that the 
nonlinear terms identically vanish, one obtains evolution equations for
$\Psi_\alpha$ and $\Psi_{2\alpha}$,
\begin{eqnarray}
\label{Eevolution}
\frac{\d}{\dt}\Psi_\alpha&=&-2\nu_\mu\Psi_{2\mu+\alpha}
+\langle\psi f\rangle,\\
\label{Zevolution}
\frac{\d}{\dt}\Psi_{2\alpha}&=&-2\nu_\mu\Psi_{2\mu+2\alpha}
+\langle(-\Delta)^{\alpha/2}\psi f\rangle.
\end{eqnarray}
Using the Cauchy--Schwarz and Young inequalities and because the 
forcing spectral support is confined to the wavenumber interval 
$[k_1,k_2]$, one obtains the upper bounds on the injection terms 
in (\ref{Eevolution}) and (\ref{Zevolution}):
\begin{eqnarray}
\label{forcebounds}
\langle\psi f\rangle&\le&\langle|(-\Delta)^{\mu/2+\alpha/4}\psi|^2\rangle^
{1/2}\langle|(-\Delta)^{-\mu/2-\alpha/4}f|^2\rangle^{1/2}\nonumber\\
&\le&\nu_\mu\Psi_{2\mu+\alpha}+\frac{1}{2\nu_\mu}
\langle|(-\Delta)^{-\mu/2-\alpha/4}f|^2\rangle\nonumber\\
&\le&\nu_\mu\Psi_{2\mu+\alpha}+\frac{\langle|f|^2\rangle}
{2\nu_\mu k_1^{2\mu+\alpha}},\nonumber\\
\langle(-\Delta)^{\alpha/2}\psi f\rangle&\le&
\langle|(-\Delta)^{\mu/2+\alpha/2}\psi|^2\rangle^{1/2}
\langle|(-\Delta)^{-\mu/2}f|^2\rangle^{1/2}\nonumber\\
&\le&\nu_\mu\Psi_{2\mu+2\alpha}+\frac{1}{2\nu_\mu}
\langle|(-\Delta)^{-\mu/2}f|^2\rangle\nonumber\\
&\le&\nu_\mu\Psi_{2\mu+2\alpha}
+\frac{\langle|f|^2\rangle}{2\nu_\mu k_1^{2\mu}}.
\end{eqnarray}
The substitution of (\ref{forcebounds}) in (\ref{Eevolution}) and 
(\ref{Zevolution}) yields
\begin{eqnarray}
\label{evolbounds}
\frac{\d}{\dt}\Psi_\alpha&\le&-\nu_\mu\Psi_{2\mu+\alpha}
+\frac{\langle|f|^2\rangle}{2\nu_\mu k_1^{2\mu+\alpha}},\nonumber\\
\frac{\d}{\dt}\Psi_{2\alpha}&\le&-\nu_\mu\Psi_{2\mu+2\alpha}
+\frac{\langle|f|^2\rangle}{2\nu_\mu k_1^{2\mu}}.
\end{eqnarray}
These inequalities readily admit the upper bounds on the asymptotic 
averages of $\Psi_{2\mu+\alpha}$ and $\Psi_{2\mu+2\alpha}$:
\begin{eqnarray}
\label{evolbound1}
\limsup_{t\rightarrow\infty}\frac{1}{t}\int_0^t\Psi_{2\mu+\alpha}\,\d\tau
&\le&\frac{1}{2\nu_\mu^2k_1^{2\mu+\alpha}}
\limsup_{t\rightarrow\infty}\frac{1}{t}\int_0^t\langle|f|^2\rangle\,\d\tau,\\
\label{evolbound2}
\limsup_{t\rightarrow\infty}\frac{1}{t}\int_0^t\Psi_{2\mu+2\alpha}\,\d\tau
&\le&\frac{1}{2\nu_\mu^2k_1^{2\mu}}
\limsup_{t\rightarrow\infty}\frac{1}{t}\int_0^t\langle|f|^2\rangle\,\d\tau.
\end{eqnarray}
For a time-independent forcing $f$, the asymptotic average on the 
right-hand sides of (\ref{evolbound1}) and (\ref{evolbound2}) reduces
to $\langle|f|^2\rangle$.

The upper bound (\ref{evolbound1}) on the asymptotic average of 
$\Psi_{2\mu+\alpha}$ is interesting, particularly for the unbounded 
case. This bound prohibits the spectrum $\Psi_{2\mu+\alpha}(k)$ 
[$\Psi_\alpha(k)$] from acquiring a spectral scaling not shallower 
than $k^{-1}$ [$k^{-2\mu-1}$] in the limit $k\rightarrow0$ because 
otherwise, a divergence of $\Psi_{2\mu+\alpha}$ would entail. This 
means that a constant inverse flux of $\Psi_\alpha$ (if it is realizable) 
ought to proceed toward $k=0$ (in the limit $t\rightarrow\infty$) 
via a spectrum $\Psi_\alpha(k)$ shallower than $k^{-2\mu-1}$. This 
observation is intuitively obvious from the physical point of view: 
the proceeding of a sustainable inverse cascade is incompatible with 
a persistent increase in the dissipation of the inverse-cascading 
quantity. Note that the critical scaling 
$\Psi_\alpha(k)\propto k^{-2\mu-1}$ (for $k<k_1$) may be realizable 
only down to a finite wavenumber $k_0>0$ and necessarily corresponds 
to an asymptotic absolute equilibrium of $\Psi_\alpha$.

For the NS system ($\alpha=2$ and $\mu=1$), 
$\Psi_{2\mu+\alpha}=\Psi_4$ (the enstrophy) is the dissipation agent 
of the energy $\Psi_2$. The above analysis implies that the enstrophy 
(energy) spectrum in the inverse-cascading region is no steeper than 
$k^{-1}$ ($k^{-3}$). This constraint turns out to be ``generous'', as 
the Kolmogorov--Kraichnan $k^{-5/3}$ energy spectrum is much shallower 
than $k^{-3}$. For SQG turbulence ($\alpha=1$ and $\mu=1/2$), the 
energy $\Psi_2$ is the dissipation agent of the inverse-cascading 
candidate $\Psi_1$. Hence, the energy asymptotic 
average is bounded. This condition requires that $\Psi_2(k)$ in the 
low-wavenumber region be no steeper than $k^{-1}$. Tran and Bowman 
\cite{Tran03b} have previously noted this requirement.

{\bf Remark 1.} Like the constraint in the NS case, the constraint 
on the spectral steepness of the low-wavenumber region for SQG 
turbulence is the threshold at which an inverse cascade is not 
sustainable. To allow for a steady positive growth rate 
of $\Psi_1$, an energy spectrum shallower than $k^{-1}$ in the 
low-wavenumber range is required. 

%{\bf Remark 2.} For NS turbulence, a persistent inverse energy 
%cascade, transferring energy to the large scales via the 
%Kolmogorov--Kraichnan $k^{-5/3}$ inertial range, is not realizable 
%for $\mu\le1/3$ because the dissipation of energy would then scale 
%as $k^{-1}$, which diverges as $k\rightarrow0$. Note that an inverse 
%cascade transferring energy to the large scales via a spectrum 
%shallower than $k^{-5/3}$ cannot be ruled out in this case.

A result derived by Tran and Shepherd \cite{Tran02a} for NS turbulence 
confined to a doubly periodic domain is now generalized to the present
problem. For simplicity, it is assumed that the forcing satisfies
$s^\alpha\langle\psi f\rangle=\langle(-\Delta)^{\alpha/2}\psi f\rangle$, 
where $s$ is some (constant) wavenumber. (The result to follow can be
readily generalized to a less restrictive forcing class \cite{Tran02a}.)
This condition can be realized in several ways. The simplest example 
\cite{Tran02a} is the case of a monoscale forcing at a wavenumber $s$,
i.e. a forcing $f$ for which $(-\Delta)^\theta f=s^{2\theta}f$. 
Another example \cite{Shepherd87,Tran03a,Tran03c,Tran02a} is
\begin{eqnarray}
\label{forcing}
\widehat{f}(\bm k)&=&\frac{\epsilon}{|K|}
\frac{\widehat\psi(\bm k)}{2\Psi(k)}, 
\end{eqnarray}
for $k\in K=[k_1,k_2]$, and $\widehat{f}(\bm k)=0$ otherwise. Here,
$\epsilon$ is a positive number, and $|K|$ is the width (the number of
discrete wavenumbers) of the forcing band for the continuous (discrete)
case. For this forcing, the spectral injection rate of 
$\Psi_\alpha(k)$ is $F(k)=\epsilon/|K|$. Hence, the respective 
injection rates of $\Psi_\alpha$ and $\Psi_{2\alpha}$ are given by
\begin{eqnarray}
\int_KF(k)\,\dk&=&\epsilon,\\
\int_Kk^\alpha F(k)\,\dk&=&\frac{\epsilon}{|K|}
\int_Kk^\alpha\,\dk=s^\alpha\epsilon,
\end{eqnarray}
where $s^\alpha$ is the mean of $k^\alpha$ over the forcing band $K$.
This forcing is used in the numerical simulations of Section 7, where
$\epsilon$ is a constant, yielding constant injection rates.

Multiplying (\ref{Eevolution}) by $s^\alpha$ and subtracting the 
resulting equation from (\ref{Zevolution}), noting that the forcing
terms cancel, one obtains
\begin{eqnarray}
\label{dynconstraint}
\frac{\d}{\dt}(\Psi_{2\alpha}-s^\alpha\Psi_\alpha)
&=&2\nu_\mu(s^\alpha\Psi_{2\mu+\alpha}-\Psi_{2\mu+2\alpha}).
\end{eqnarray}
The difference on the right-hand side can be written as
\begin{eqnarray}
s^\alpha\Psi_{2\mu+\alpha}-\Psi_{2\mu+2\alpha}
&=&-s^{2\mu}(\Psi_{2\alpha}-s^\alpha\Psi_\alpha)
-\int_0^\infty(s^{2\mu}-k^{2\mu})(s^\alpha-k^\alpha)\Psi_\alpha(k)\,\dk.
\nonumber
\end{eqnarray}
Substituting this expression into (\ref{dynconstraint}) yields
\begin{eqnarray}
\label{dynconstraint1}
\frac{\d}{\dt}(\Psi_{2\alpha}-s^\alpha\Psi_\alpha)
&=&-2\nu_\mu s^{2\mu}(\Psi_{2\alpha}-s^\alpha\Psi_\alpha)
-2\nu_\mu\int_0^\infty(s^{2\mu}-k^{2\mu})(s^\alpha-k^\alpha)
\Psi_\alpha(k)\,\dk.\nonumber\\
\end{eqnarray}
For $\mu\ge0$ and $\alpha>0$, the integral in
(\ref{dynconstraint1}) is non-negative. The first order equation 
(\ref{dynconstraint1}) is dissipative and forced by a non-positive 
term. It follows that $\Psi_{2\alpha}-s^\alpha\Psi_\alpha$ becomes
non-positive as $t\rightarrow\infty$. Hence, the constraint
\begin{eqnarray}
\label{dynconstraint2} 
\Psi_{2\alpha}&\le&s^\alpha\Psi_\alpha
\end{eqnarray}
holds asymptotically. 

{\bf Remark 2.} Actually, the constraint (\ref{dynconstraint2}) holds 
for all $t\ge T$, where $T$ is a finite time determined by the initial 
condition, except for some trivial cases, which are rather irrelevant
in the context of turbulence (see \cite{Tran02a}). This constraint is 
mathematically interesting in the bounded case, as the inequality is 
in the opposite sense to Sobolev-type inequalities. In an unbounded 
domain where $\Psi_\alpha$ may grow unbounded whether or not an 
inverse cascade exists, (\ref{dynconstraint2}) becomes obvious. 

{\bf Remark 3.} The dynamical constraint (\ref{dynconstraint2}) 
holds if the single viscous operator is replaced by a combination
$\sum_\mu\nu_\mu(-\Delta)^\mu$ for $\nu_\mu,\mu\ge0$, where the 
summation is taken over all dissipation channels involved 
\cite{Tran02a}. It also holds, with $s$ replaced by $k_2$, for a 
forcing satisfying $k_1^\alpha\int_KF(k)\,\dk\le\int_Kk^\alpha 
F(k)\,\dk\le k_2^\alpha\int_KF(k)\,\dk$. This double inequality 
automatically holds for any non-negative spectral injection, i.e. 
$F(k)\ge 0$ for all $k\in K$.

\section{Conservative transfer}

This section examines the nonlinear transfer of the quadratic 
invariants and their dynamical behaviours, particularly the 
possibility of a direct cascade of $\Psi_{2\alpha}$. A 
Fj{\o}rtoft-type analysis \cite{Fjortoft53} concerning the 
collective transfer of all nonlinearly interacting triads is 
presented. The simple technique of Tran and Bowman \cite{Tran03c} 
is employed in deriving a constraint on the spectral slope of the 
supposed direct-cascading range in terms of the growth rate of the 
inverse-cascading candidate and the dissipation wavenumber.
These analyses finally lead to a self-consistent dynamical picture, 
which exhibits no direct cascade.

\subsection{Inverse vs. direct transfer}
Due to the conservation laws (\ref{conservation1}), the nonlinear 
transfer of $\Psi_\alpha$ and $\Psi_{2\alpha}$ obeys certain restrictions.
Suppose that an initial $\Psi_\alpha^0$ at wavenumber $s$, which 
corresponds to an initial $\Psi_{2\alpha}^0=s^\alpha\Psi_\alpha^0$, is
nonlinearly redistributed in wavenumber space. Because of the 
conservation of $\Psi_{2\alpha}$, the redistribution of an amount 
$\Psi^+_\alpha$ beyond some given wavenumber $k_+>s$ must satisfy 
$k_+^\alpha\Psi^+_\alpha<s^\alpha\Psi_\alpha^0$, or equivalently, 
$\Psi^+_\alpha/\Psi_\alpha^0<(s/k_+)^\alpha$. Hence, no significant 
fractions of $\Psi_\alpha^0$ are allowed to be transferred toward 
wavenumbers $k\gg s$. This condition corresponds to the well-known 
prohibition of a direct cascade of energy in the NS case. The 
constraint is more severe for a larger $\alpha$ in the sense that a 
larger $\alpha$ implies a smaller permissible $\Psi^+_\alpha$ to be 
redistributed beyond a given $k_+>s$. Similarly, due to the conservation 
of $\Psi_\alpha$, the redistribution of an amount $s^\alpha\Psi^-_\alpha$ 
to the wavenumbers lower than some given wavenumber $k_-<s$ must satisfy 
$s^\alpha\Psi^-_\alpha/k_-^\alpha<\Psi^0_\alpha$, or equivalently, 
$s^\alpha\Psi^-_\alpha/(s^\alpha\Psi^0_\alpha)<(k_-/s)^\alpha$. 
Hence, no significant fractions of $s^\alpha\Psi^0_\alpha$ are allowed 
to be transferred toward wavenumbers $k\ll s$. On the other hand, 
the redistribution of any amount $\Psi'_\alpha<\Psi^0_\alpha$ toward 
the vanishing wavenumbers is permitted. Such a transfer induces a 
loss of $s^\alpha\Psi'_\alpha$ in $\Psi_{2\alpha}$. The conservation 
of $\Psi_{2\alpha}$ then requires that the remainder 
$\Psi_\alpha^0-\Psi'_\alpha$ be redistributed toward the high 
wavenumbers, to make up for the lost amount $s^\alpha\Psi'_\alpha$. 
For a non-negligible remainder $\Psi_\alpha^0-\Psi'_\alpha$ and large
$\alpha$, the lost amount $s^\alpha\Psi'_\alpha$ can be recovered with
a redistribution of $\Psi_\alpha^0-\Psi'_\alpha$ toward wavenumbers
slightly larger than $s$, resulting in no direct cascade. 
This means that a direct cascade is possible only in the limit 
$\Psi'_\alpha\rightarrow\Psi_\alpha^0$. In other words, an inverse 
cascade can exist in the complete absence of a direct cascade, and 
the latter is not permitted until virtually all of $\Psi_\alpha$ 
cascades to the vanishing wavenumbers. This observation is consistent 
with the results from numerical simulations of 2D NS turbulence that 
have found an inverse energy cascade, regardless of what happened to 
the enstrophy \cite{Boffetta00,Borue94,Frisch84,Tran03c}.

\subsection{Constraint on spectral slope for $k>s$}
A more quantitative version of the preceding analysis is in order. 
The dual-cascade scenario is now examined in the framework 
of the forced-dissipative picture. A constraint on the spectral slope 
of the supposed direct-cascading range is derived in terms of the 
growth rate of $\Psi_\alpha$ and the dissipation wavenumber. 

The evolution of $\Psi_\alpha$ and $\Psi_{2\alpha}$ is governed by
\begin{eqnarray}
\label{Eevolution1}
\frac{\d}{\dt}\Psi_\alpha&=&-2\nu_\mu\Psi_{2\mu+\alpha}
+\int_KF(k)\,\dk,\\
\label{Zevolution1}
\frac{\d}{\dt}\Psi_{2\alpha}&=&-2\nu_\mu\Psi_{2\mu+2\alpha}
+\int_Kk^\alpha F(k)\,\dk.
\end{eqnarray}  
Let us consider quasi-steady dynamics, where a steady spectrum has been 
established down to a wavenumber $k_0\ll s$. The injections have become 
steady and satisfy $s^\alpha\int_KF(k)\,\dk=\int_Kk^\alpha F(k)\,\dk$, 
where $s$ is a wavenumber in the forcing region. The evolution of 
$\Psi_{2\alpha}$ has reached equilibrium, while $\Psi_\alpha$ continues 
to cascade toward wavenumbers $k<k_0$ at a steady growth rate 
$\d\Psi_\alpha/\dt=\epsilon_0$.\footnote{It should be emphasized that
although evidence exists of an inverse cascade for various values
of $\alpha$, one cannot be sure that an inverse cascade exists for 
all $\alpha>0$. Section 6 deals with the non-cascading case 
$\epsilon_0=0$, which is applicable to both bounded and unbounded 
systems in equilibrium.}
It follows from (\ref{Eevolution1}) and (\ref{Zevolution1}) that
\begin{eqnarray}
\label{P/Z}
\frac{\Psi_{2\mu+2\alpha}}{\Psi_{2\mu+\alpha}}
&=&s^\alpha\frac{\epsilon}{\epsilon-\epsilon_0},
\end{eqnarray}
where $\epsilon=\int_KF(k)\,\dk$. For simplicity, it is assumed 
that the quasi-steady spectrum $\Psi_\alpha(k)$ can be approximated by 
\cite{Tran03a}
\begin{eqnarray}
\label{spectrum}
\Psi_\alpha(k) &=& \cases{
ak^{-\gamma}&if $k_0<k<s$,\cr
bk^{-\beta}&if $s<k<k_\nu$,\cr}
\end{eqnarray}
where $a,~b,~\gamma,~\beta$ are constants, and $k_\nu$ is the
highest wavenumber in the range $k^{-\beta}$, beyond which the spectrum
is supposed to be steeper than $k^{-\beta}$. The ratio on the left-hand 
side of (\ref{P/Z}) can be estimated as
\begin{eqnarray}
\label{spectrum1}
\frac{\Psi_{2\mu+2\alpha}}{\Psi_{2\mu+\alpha}}
&\ge&\frac{a\int_{k_0}^s k^{2\mu+\alpha-\gamma}\,\dk
+b\int_s^{k_\nu}k^{2\mu+\alpha-\beta}\,\dk}
{a\int_{k_0}^s k^{2\mu-\gamma}\,\dk
+b\int_s^{\infty}k^{2\mu-\beta}\,\dk}\nonumber\\
&=&s^{\alpha}\frac{\int_{k_0/s}^1\kappa^{2\mu+\alpha-\gamma}\,\d\kappa
+\int_{s/k_\nu}^1\kappa^{\beta-2\mu-\alpha}\,\d\kappa}
{\int_{k_0/s}^1\kappa^{2\mu-\gamma}\,\d\kappa
+\int_0^1\kappa^{\beta-2\mu}\,\d\kappa}. 
\end{eqnarray}
The inequality results from dropping from the estimate of 
$\Psi_{2\mu+2\alpha}$ the spectral contribution beyond $k_\nu$ and 
extending the range $bk^{-\beta}$ to infinity in the estimate of 
$\Psi_{2\mu+\alpha}$. The equality is obtained by making the 
respective substitutions $\kappa=k/s$ and $\kappa=s/k$ in the two 
integrals in each of the numerator and denominator plus the 
continuity relation $as^{-\gamma}=bs^{-\beta}$. It follows that
\begin{eqnarray}
\label{P/Z1}
\frac{\int_{s/k_\nu}^1\kappa^{\beta-2\mu-\alpha-2}\,\d\kappa}
{\int_{k_0/s}^1\kappa^{2\mu-\gamma}\,\d\kappa
+\int_0^1\kappa^{\beta-2\mu-2}\,\d\kappa} 
&\le&\frac{\epsilon}{\epsilon-\epsilon_0}.
\end{eqnarray}

A direct cascade of $\Psi_{2\alpha}$ requires that its dissipation 
occurs primarily in the small scales. Accordingly, the spectrum
$\Psi_{2\mu+2\alpha}(k)$ must be shallower than $k^{-1}$. It is 
interesting to consider the regime where the onset of direct-cascading 
dynamics is possible, i.e., the regime for which $\Psi_{2\mu+2\alpha}(k)$ 
approximately scales as $k^{-1}$ for $k>s$. This scaling means 
$-\beta+2\mu+\alpha\approx-1$, or equivalently $\beta-2\mu-2\approx\alpha-1$. 
The second integral in the denominator on the left-hand side of 
(\ref{P/Z1}) is then of order unity (or smaller if $\alpha\gg1$). Now
it is seen in Section 3 that a persistent inverse cascade
($\epsilon_0>0$) requires that $\Psi_{2\mu+\alpha}(k)$ be shallower 
than $k^{-1}$ in the inverse-cascading range, i.e. $2\mu-\gamma>-1$. 
This requirement implies that the first integral in the denominator on the 
left-hand side of (\ref{P/Z1}) is of order unity. Hence, the denominator 
on the left-hand side of (\ref{P/Z1}) is of order unity. On the other 
hand, the integral in the numerator can be significantly larger than 
unity. In particular, this integral approaches the value $\ln(k_\nu/s)$
in the limit $\beta-2\mu-\alpha-1\rightarrow0$ and can be large for 
a large ratio $k_\nu/s$. This result implies that the left-hand side 
of (\ref{P/Z1}) can be significantly larger than unity in the 
non-direct-cascading regime. The constraint (\ref{P/Z1}) then requires 
that $\epsilon_0$ be sufficiently close to $\epsilon$. Hence, an inverse 
cascade with various strength (various values of the ratio 
$\epsilon_0/\epsilon$) can exist in the absence of a direct cascade, 
as was previously argued in this section without an explicit dissipation 
mechanism. Finally, in the limit $s/k_\nu\rightarrow0$, as is commonly 
assumed for high-Reynolds number turbulence, a direct cascade, 
regardless of the spectral slope $-\beta$, requires the limit
$\epsilon_0\rightarrow\epsilon$.

{\bf Remark 4.} For NS turbulence, the critical slope of the energy 
spectrum, for $k>s$, at the onset of a possible direct cascade is 
$-5$. The KLB direct-enstrophy-cascading range scales as $k^{-3}$, 
which is much shallower than the critical scaling $k^{-5}$. (The 
spectral discrepancy between these two scaling laws is huge.) 
$\beta$ has been proposed to have some favourable values between 
$3$ and $5$, according to the theories of Moffatt \cite{Moffatt86} 
($\beta=11/3$), Saffman \cite{Saffman71} ($\beta=4$), and Sulem and 
Frisch \cite{Sulem75} ($\beta\le11/3$). For SQG turbulence, the 
critical slope is $-2$. The Kolmogorov--Kraichnan 
direct-energy-cascading range scales as $k^{-5/3}$, which is 
slightly shallower than the critical scaling $k^{-2}$.

The question remains as to whether a direct cascade is realizable in 
unbounded fluids. The present analysis narrows down the possibility 
of a direct cascade to the regime in which the growth rate of 
$\Psi_\alpha$ approaches its own injection rate. If a direct cascade 
could be realizable and if a spectrum much shallower than the critical 
spectrum could be achieved (e.g., a $k^{-3}$ scaling in NS turbulence), 
the critical scaling would intuitively be realizable at relatively 
moderate Reynolds numbers, and the direct-cascading range would become 
shallower as the Reynolds number increases. Numerical simulations may 
be able to probe into this critical regime and help assess the 
existence or non-existence of a direct cascade. Given the low 
resolutions available in current computers, a plausible possibility 
is to explore how $\epsilon_0$, $\beta$, and $k_\nu$ adjust with 
respect to the Reynolds number in the non-direct-cascading regime 
\cite{Tran03c}. This information may then be extrapolated to the 
critical limit. A difficulty with this scheme is that one might not 
be able to get sufficiently close to the critical regime to make the 
extrapolation meaningful.\footnote{For 2D NS turbulence, one would 
need several wavenumber decades to simulate an inverse energy cascade 
that carries about 90\% of the injected energy to the large scales 
\cite{Tran03c}. [This requirement can be readily deduced from 
(\ref{dissipationbound}).] This resolution is already a challenge for 
current computers. In fact, no numerical simulations performed thus 
far have achieved a resolution better than four wavenumber decades.} 

\subsection{A non-direct-cascading dynamical picture}
It is interesting to consider a self-consistent dynamical picture, 
which is in accord with the preceding analysis and exhibits no direct 
cascade, even in the limit of infinite Reynolds number. This discussion 
is concerned with the plausibility of a direct cascade rather than 
with an attempt to find an alternative to the dual-cascade theory. 
In the limit $\nu_\mu\rightarrow0$, the equilibrium of $\Psi_{2\alpha}$ 
implies that $\Psi_{2\mu+2\alpha}$ diverges as $\nu_\mu^{-1}$. In the 
absence of a direct cascade, this condition can be accounted for by 
both the logarithmic divergence of $\ln(k_\nu/s)$, as 
$k_\nu\rightarrow\infty$, and an increase of $b$. 
The latter factor is relevant since it is not expected that $k_\nu$ 
grows exponentially as $\nu_\mu\rightarrow0$. In any case, one has 
$s^\alpha\epsilon/2\nu_\mu=\Psi_{2\mu+2\alpha}\ge b\ln(k_\nu/s)$.
Meanwhile, $\Psi_{2\mu+\alpha}$ is given by
\begin{eqnarray}
\Psi_{2\mu+\alpha}&=&bs^{-\alpha}\left(\frac{1}{2\mu+1-\gamma}
+\frac{1}{\alpha}\right),
\end{eqnarray}
for $2\mu+1-\gamma>0$. The dissipation of $\Psi_\alpha$ can be bounded
from above by
\begin{eqnarray}
\label{dissipationbound}
2\nu_\mu\Psi_{2\mu+\alpha}&\le&\frac{\epsilon}{\ln(k_\nu/s)}
\left(\frac{1}{2\mu+1-\gamma}+\frac{1}{\alpha}\right).
\end{eqnarray}

This result implies that $2\nu_\mu\Psi_{2\mu+\alpha}\rightarrow0$ as 
$k_\nu\rightarrow\infty$ ($\nu_\mu\rightarrow0$). Hence, in the limit 
$\nu_\mu\rightarrow0$, no direct cascade would be needed in order for 
an inverse cascade to proceed toward $k=0$, carrying with it virtually 
all of the input of $\Psi_\alpha$. Thus, this hypothetical picture 
depicts non-direct-cascading dynamics for all Reynolds numbers. As 
$\nu_\mu$ becomes ever smaller, the range $bk^{-\beta}$ gets ever wider, 
$\beta$ gets ever closer to the critical value $\beta=2\mu+\alpha+1$, 
and $b$ increases somewhat less rapidly than $\nu_\mu^{-1}$. Eventually, 
$\Psi_\theta$ (for all real $\theta$) diverges as $\nu_\mu\rightarrow0$. 
This feature is completely absent from the dual-cascade picture,\footnote
{In the dual-cascade picture, $b$ is essentially insensitive to 
$\nu_\mu$ in the limit $\nu_\mu\rightarrow0$. The divergence of 
$\Psi_{2\mu+2\alpha}$, as $\nu_\mu\rightarrow0$, is mainly due to the 
increase of $k_\nu$. $\Psi_{2\mu+\alpha}$ (and possibly other quadratic 
quantities) remains finite in the limit $\nu_\mu\rightarrow0$. 
Therefore, the dual-cascade picture poses a ``sudden'' blow-up at 
$\nu_\mu=0$. This behaviour is somewhat harder to perceive than the 
``smooth'' blow-up in the non-direct-cascading picture.} where 
$\Psi_{2\mu+\alpha}$ is supposed to remain finite in the limit 
$\nu_\mu\rightarrow0$. Note that the present discussion would apply 
to both bounded and unbounded systems. The only essential difference 
between these systems would be that for the bounded system, the 
inverse cascade is halted at the largest available scale, upon which 
significant departures in the dynamics of the two systems take place. 
The approach to equilibrium after the arrest of the inverse cascade 
in 2D NS turbulence was numerically studied by Borue \cite{Borue94}. 
Also, a physical interpretation of the observed $k^{-3}$ spectrum 
at the large scales was given. Tran and Bowman \cite{Tran03a,Tran03c}
observed and explained a similar result. 

The existence of a weak inverse energy cascade in the complete absence 
of a direct enstrophy cascade for moderate-Reynolds number NS turbulence 
is strongly supported by the theoretical arguments and numerical results 
of Tran and Bowman \cite{Tran03a,Tran03c} and by the numerous numerical 
investigations targeted at the inverse cascade in the literature.\footnote
{Those investigations inadvertently do not recognize the absence of a 
direct cascade for obvious reason: the forcing region is usually chosen 
as close as possible to the largest available wavenumber to obtain a wide
inverse-cascading range, thereby allowing for no simultaneous 
observations of the enstrophy dynamics.} Section 7 provides further
numerical evidence for the realizability of non-direct-cascading dynamics.
The results reported there also show the formation of an enstrophy range
shallower than $k^{-5}$ during the transient dynamics, thereby suggesting
that the turbulence would become non-direct-cascading by evolving through
a direct-cascading stage.

\section{Non-conservative transfer}
The creation and depletion of quadratic quantities other than the two
invariants by nonlinear transfer are now investigated. For an arbitrary 
spectrum, subject to the finiteness of the quadratic quantities under 
consideration, one has the following interpolation-type inequality, 
which can be shown by H\"{o}lder's inequality:
\begin{eqnarray}
\label{Holder}
\Psi_{\theta_2+\theta_3}^{\theta_1}&\le&
\Psi_{\theta_1+\theta_3}^{\theta_2}\Psi_{\theta_3}^{\theta_1-\theta_2},
\end{eqnarray}
for $\theta_1>\theta_2>0$ and $\theta_3$ is an arbitrary real number. 
This inequality is geometry-independent, so that it applies to both 
bounded and unbounded cases. The equality occurs if and only if 
$\Psi(k)\neq0$ is monoscale, i.e., if $\Psi(k)$ is nonzero only at 
a single wavenumber (a delta function for the case of continuous 
spectrum). This trivial case is excluded so that only the strict 
version of (\ref{Holder}) is applied in what is to follow. By applying 
(\ref{Holder}) with 
$(\theta_1,\theta_2,\theta_3)=(\alpha,\theta-\alpha,\alpha)$
one obtains
\begin{eqnarray}
\label{Holder1}
\Psi_\theta&<&\Psi_{2\alpha}^{\theta/\alpha-1}\Psi_\alpha^
{2-\theta/\alpha}\mbox{~~~~for~~~~}\theta\in(\alpha,2\alpha).
\end{eqnarray}
Similarly, the two substitutions 
$(\theta_1,\theta_2,\theta_3)=(\theta-\alpha,\alpha,\alpha)$ and
$(\theta_1,\theta_2,\theta_3)=(2\alpha-\theta,\alpha-\theta,\theta)$
into (\ref{Holder}) yield
\begin{eqnarray}
\label{Holder2}
\Psi_\theta&>&\Psi_{2\alpha}^{\theta/\alpha-1}
\Psi_\alpha^{2-\theta/\alpha}\mbox{~~~~for~~~~}\theta\not\in[\alpha,2\alpha].
\end{eqnarray}
The quantities on the right-hand sides of (\ref{Holder1}) and 
(\ref{Holder2}) are conserved by nonlinear transfer (because $\Psi_\alpha$
and $\Psi_{2\alpha}$ are invariant) and equal to $\Psi_\theta^0$, for an 
initial monoscale $\Psi_\theta^0$. Therefore, 
the subsequent nonlinear transfer necessarily results in the depletion 
(creation) of $\Psi_\theta$ for $\theta\in(\alpha,2\alpha)$ 
($\theta\not\in[\alpha,2\alpha]$). It should be noted that continuous 
increase (decrease) of the non-conserved quadratic quantities from 
an initial monoscale spectrum is possible but cannot be assured, due
to the reversible nature of the inviscid problem.

The above picture, when applied to forced-dissipative systems in 
equilibrium or in quasi-steady dynamics, implies that the nonlinear 
transfer destroys (enhances) the injection of $\Psi_\theta$ from a
monoscale forcing for $\theta\in(\alpha,2\alpha)$ 
($\theta\not\in[\alpha,2\alpha]$). In other words, the integrated
transfer $\int_0^\infty k^{\theta-\alpha}T_\alpha(k)\,\dk$ in the
evolution equation
\begin{eqnarray}
\frac{\d}{\dt}\Psi_\theta&=&\int_0^\infty k^{\theta-\alpha}
T_\alpha(k)\,\dk-2\mu\Psi_{2\mu+\theta}+\int_Kk^{\theta-\alpha}F(k)\,\dk
\end{eqnarray}
is negative (positive) for $\theta\in(\alpha,2\alpha)$ 
($\theta\not\in[\alpha,2\alpha]$). To see this result explicitly, note 
that in equilibrium or quasi-steady dynamics, $T_\alpha(k)\ge0$ for 
$k\neq s$ because $F(k)=0$ for $k\neq s$, and the dissipation is
non-negative [cf. (\ref{Psi(k)evolution})]. Similar to (\ref{Holder}),
the non-negative (and non-monoscale) transfer function 
$T_\alpha(k)$, for $k\neq s$, satisfies 
\begin{eqnarray}
\label{Holder3}
\left(\int_{k\neq s}k^{\theta_2+\theta_3}T_\alpha(k)\,\dk\right)^{\theta_1}
&<&
\left(\int_{k\neq s}k^{\theta_1+\theta_3}T_\alpha(k)\,\dk\right)^{\theta_2}
\left(\int_{k\neq s}k^{\theta_3}T_\alpha(k)\,\dk\right)^{\theta_1-\theta_2}
\end{eqnarray}
for $\theta_1>\theta_2>0$ and $\theta_3$ is an arbitrary real number.
Upon substituting 
$(\theta_1,\theta_2,\theta_3)=(\alpha,\theta-\alpha,0)$ into 
(\ref{Holder3}) and using the conservation laws (\ref{conservation1}), 
one finds
\begin{eqnarray}
\int_{k\neq s}k^{\theta-\alpha}T_\alpha(k)\,\dk
&<&
\left(\int_{k\neq s}k^\alpha T_\alpha(k)\,\dk\right)^{\theta/\alpha-1}
\left(\int_{k\neq s}T_\alpha(k)\,\dk\right)^{2-\theta/\alpha}\nonumber\\
&=&\left(s^\alpha|T_\alpha(s)|\right)^{\theta/\alpha-1}
|T_\alpha(s)|^{2-\theta/\alpha}=s^{\theta-\alpha}|T_\alpha(s)|
\end{eqnarray}
for $\theta\in(\alpha,2\alpha)$, or equivalently,
\begin{eqnarray}
\int_0^\infty k^{\theta-\alpha}T_\alpha(k)\,\dk&<&0.
\end{eqnarray}
This result means that for $\theta\in(\alpha,2\alpha)$, $\Psi_\theta$ 
is destroyed by nonlinear transfer. Similarly, with the substitutions 
$(\theta_1,\theta_2,\theta_3)=(\theta-\alpha,\alpha,0)$ and
$(\theta_1,\theta_2,\theta_3)=(2\alpha-\theta,\alpha-\theta,\theta-\alpha)$, 
one obtains
\begin{eqnarray}
\int_0^\infty k^{\theta-\alpha}T_\alpha(k)\,\dk&>&0,
\end{eqnarray} 
for $\theta\not\in[\alpha,2\alpha]$. This result means that for 
$\theta\not\in[\alpha,2\alpha]$, $\Psi_\theta$ is created by
nonlinear transfer.

The creation and depletion of $\Psi_\theta$ for different values of 
$\alpha$ are interesting; here, only some typical cases are considered.
For $\alpha<2$, the enstrophy $\Psi_4$ (and vorticity) can be created. 
This enstrophy creation arguably occurs in SQG turbulence ($\alpha=1$), 
as its inviscid unforced dynamics is known to resemble that of the 
three-dimensional Euler equation in many aspects, particularly in the 
possibility of spontaneous development of singularities 
\cite{Constantin94a,Constantin94b}. Note that the enstrophy creation
need not be a consequence of a direct cascade and that the created 
enstrophy is subject to viscous dissipation in dissipative 
systems. For $2<\alpha<4$ such as in RSF turbulence ($\alpha=3$), 
the energy is created while the enstrophy is destroyed. The creation 
of energy occurs toward the low wavenumbers and evades viscous 
dissipation. If an inverse cascade of $\Psi_\alpha$ proceeds toward 
$k=0$ at a steady growth rate, the energy growth rate increases 
without bound. When $\alpha>4$, both the energy and enstrophy are 
created, and this creation occurs toward the low wavenumbers. Both 
the energy and enstrophy growth rates increase without bound in the 
presence of a steady growth rate of $\Psi_\alpha$. When $\alpha<1$, 
both the energy and enstrophy are created, and this creation occurs 
toward the high wavenumbers. The NS system happens to have the critical 
value $\alpha=2$, for which both the energy and enstrophy are conserved 
by advective nonlinearities. 

\section{Constraints on the slopes of equilibrium spectrum}

For fluids in equilibrium, $\epsilon_0=0$ and (\ref{P/Z}) reduces 
to the balance equation $s^\alpha\Psi_{2\mu+\alpha}=\Psi_{2\mu+2\alpha}$. 
Assuming the spectral scaling (\ref{spectrum}) and following the 
calculations leading to (\ref{P/Z1}), one obtains the estimate of
$\Psi_{2\mu+2\alpha}-s^\alpha\Psi_{2\mu+\alpha}$:
\begin{eqnarray}
\label{balance}
\int_{k_0}^{\infty}(k^\alpha-s^\alpha)k^{2\mu}\Psi_\alpha(k)\,\dk 
&\ge&a\int_{k_0}^s(k^\alpha-s^\alpha)k^{2\mu-\gamma}\,\dk+
b\int_s^{k_\nu}(k^\alpha-s^\alpha)k^{2\mu-\beta}\,\dk\\ 
&=&as^{2\mu+\alpha+1-\gamma}\int_{k_0/s}^1(\kappa^\alpha-1)
\kappa^{2\mu-\gamma}\,\d\kappa\nonumber\\
&&+bs^{2\mu+\alpha+1-\beta}\int_{s/k_\nu}^1(1-\kappa^\alpha)
\kappa^{\beta-2\mu-\alpha-2}\,\d\kappa\nonumber\\
&=&as^{2\mu+\alpha+1-\gamma}\left(-\int_{k_0/s}^1(1-\kappa^\alpha)
\kappa^{2\mu-\gamma}\,\d\kappa+\int_{s/k_\nu}^1(1-\kappa^\alpha)
\kappa^{\beta-2\mu-\alpha-2}\,\d\kappa\right),\nonumber
\end{eqnarray} 
where the inequality results from dropping the spectral contribution
beyond~$k_\nu$, the second line is obtained by the respective changes 
of variables $\kappa=k/s$ and $\kappa=s/k$ in the two integrals on the
right-hand-side of the first line, and the third line is obtained by 
using the continuity relation $as^{-\gamma}=bs^{-\beta}$. Here, $k_0$ 
is the lowest wavenumber in the bounded case or a cutoff wavenumber 
in the unbounded case. Since the left-hand side of (\ref{balance}) 
vanishes, it can be deduced that
\begin{eqnarray}
\label{balance1}
\int_{k_0/s}^1(1-\kappa^\alpha)\kappa^{2\mu-\gamma}\,\d\kappa
&\ge&\int_{s/k_\nu}^1(1-\kappa^\alpha)
\kappa^{\beta-2\mu-\alpha-2}\,\d\kappa.
\end{eqnarray} 
For bounded high-Reynolds number turbulence driven at a relatively low 
wavenumber $s$, it is reasonable to assume that $k_0/s\ge s/k_\nu$. 
This condition requires $2\mu-\gamma\le\beta-2\mu-\alpha-2$, as the 
integrals in (\ref{balance1}) decrease if the corresponding powers of 
$\kappa$ ($2\mu-\gamma$ and $\beta-2\mu-\alpha-2$) increase. It 
follows that $\gamma+\beta\ge4\mu+\alpha+2$. 

On the other hand, the convergence of the right-hand integral in
(\ref{balance1}), as $s/k_\nu\rightarrow0$, requires $\beta>2\mu+\alpha+1$.
(Equivalently, $\Psi_{2\mu+2\alpha}(k)$ is steeper than $k^{-1}$ for $k>s$.)
That is, no direct cascade exists for equilibrium dynamics. 
The spectral contribution beyond $k_\nu$ which is dropped from 
(\ref{balance}) is then negligible and (\ref{balance1}) is 
essentially an equality. It follows that 
$\gamma+\beta\approx4\mu+\alpha+2$ if $k_0/s\approx s/k_\nu$, and 
$\gamma+\beta$ can fall slightly below $4\mu+\alpha+2$ if 
$k_0/s<s/k_\nu$. For both these cases, $\gamma\le2\mu+1$, so that 
an inverse cascade is also ruled out because the dissipation of 
$\Psi_\alpha$ cannot occur at the largest scales. Although an 
inverse cascade cannot be ruled out in the case $k_0/s>s/k_\nu$, 
such a cascade, if realizable, would only be marginal and 
characteristically different from its unbounded counterpart in 
the classical picture.

It is more informative to write the constraint 
$\gamma+\beta\ge4\mu+\alpha+2$, for $k_0/s\ge s/k_\nu$, in terms of 
the slopes of the kinetic energy spectrum $\Psi_2(k)$. Let 
$-\tilde\gamma$ and $-\tilde\beta$ denote, respectively, the slopes of 
$\Psi_2(k)$ in the ranges of wavenumbers lower and higher than the 
forcing wavenumber; $\tilde\gamma$ and $\tilde\beta$ are then related
to $\gamma$ and $\beta$ by $\tilde\gamma=\gamma+\alpha-2$ and
$\tilde\beta=\beta+\alpha-2$. It follows that 
\begin{eqnarray}
\label{slopeconstraint}
\tilde\gamma+\tilde\beta&\ge&4\mu+3\alpha-2.
\end{eqnarray}
This inequality reduces to $\tilde\gamma+\tilde\beta\ge8$ for NS 
turbulence, $\tilde\gamma+\tilde\beta\ge3$ for SQG turbulence, and 
$\tilde\gamma+\tilde\beta\ge4\mu+7$ for RSF turbulence.

Constantin \cite{Constantin02} has studied the spectral scaling of 
SQG turbulence and derived the pointwise estimate 
$\Psi_2(k)\le ck^{-2}$ in the wavenumber range $k<s$. Smith et al. 
\cite{Smith02} numerically explored this problem and found energy 
spectra considerably shallower than $k^{-1}$ in this range. The 
``energy'' spectrum plotted by Smith et al. \cite{Smith02} is 
$\Psi_1(k)$, and this spectrum appears to have a slope $\approx -1.5$. 
Hence, $\Psi_2(k)$ approximately scales\footnote {Note that Smith 
et al. \cite{Smith02} used linear drag in their simulations. Hence, 
the observed $k^{-0.5}$ scaling may in principle differ from that 
of SQG turbulence with the dissipation operator $(-\Delta)^{1/2}$. 
Nevertheless, their results provide strong evidence that the energy 
spectrum in the wavenumber range $k<s$ is shallower than $k^{-1}$.} 
as $k^{-0.5}$. Dimensional analyses predict the scaling 
$\Psi_2(k)\propto k^{-1}$ for $k<s$ \cite{Pierrehumbert94,Smith02}. 
The present estimate agrees with Smith et al.'s numerical results and
with the prediction based on dimensional analyses better than with 
Constantin's estimate. On physical grounds, a spectrum shallower than
$k^{-1}$ is more plausible, for two reasons. First, the prohibition 
of a significant inverse energy transfer in SQG turbulence (in 
marked contrast to the inverse energy cascade in NS turbulence) 
should render the energy spectrum in the wavenumber range $k<s$  
shallower than $k^{-1}$ because any spectra not shallower than 
$k^{-1}$ would require a significant inverse energy transfer. Second, 
the hypoviscous dissipation of SQG dynamics, $(-\Delta)^{1/2}$, does 
not accommodate any energy spectra not shallower than $k^{-1}$ in 
the low-wavenumber range (see Section 3). 

In a recent laboratory experiment on a rapidly rotating fluid, Baroud 
et al. \cite{Baroud02} observed a $k^{-2}$ spectrum extending for 
almost two wavenumber decades lower than the forcing wavenumber. This 
result is consistent with the constraint $\tilde\gamma+\tilde\beta\ge8$ 
obtained for the NS case, but cannot be explained within the context 
of SQG turbulence, for the reasons discussed above. A possible 
explanation for Baroud et al.'s result is that a rotating fluid, 
believed to be governed by the SQG equation,\footnote{The derivation 
of the SQG equation involves the temperature field 
\cite{Pedlosky87,Tung03}. More accurately, in order to arrive at 
the SQG equation, the vertical gradient of the temperature at the 
surface of a rotating stratified fluid is considered. Such a 
gradient is absent in Baroud et al.'s experiment since the fluid 
temperature is presumably constant throughout the rotating tank.
Hence, the results in \cite{Baroud02} cannot be attributed with 
certainty to SQG turbulence.} may still behave (to some extent) as 
a 2D NS fluid. Therefore, the dynamics of such a fluid might exhibit 
mixed characteristics of both SQG and NS turbulence. 
While a systematic analysis is required for a proper understanding 
of Baroud et al.'s experimental result, this hypothesis offers a 
possible explanation for the observed spectrum. 

\section{Numerical results}

This section presents results from numerical simulations that confirm 
the realization of an inverse energy cascade in 2D NS turbulence, where 
energy is transferred to the large scales via the Kolmogorov--Kraichnan 
spectrum $k^{-5/3}$, in the absence of a direct cascade. These results
provide numerical evidence (also see \cite{Tran03c}) for the prediction 
and motivation for the discussion in the preceding sections. For the
two sufficiently resolved Reynolds numbers, the energy growth rate
$\epsilon_0$, calculated as the inverse cascade reaches $k\approx3$, 
is larger for the higher Reynolds number. However, this preliminary 
attempt is in no position to address what happens in the high-Reynolds
number regime.

Eq. (\ref{governing}) with $\alpha=2$ and $\mu=1$ is simulated in a 
doubly periodic square of side $2\pi$. The forcing $\widehat{f}(\bm k)$ 
is given by (\ref{forcing}), where $\epsilon=1$ and $K=[49.5,50.5]$.
The enstrophy injection rate is $s^2\epsilon\approx2500$, where 
$s^2\approx2500$ is the mean of $k^2$ over $K$. The attractive aspect 
of (\ref{forcing}) is that it is steady, so that the uncertainty in
the energy transfer rate $\epsilon_0$, which is calculated according to
$\epsilon_0=\epsilon-2\nu_1\Psi_4$, is due solely to that of the enstrophy 
$\Psi_4$. It is necessary to have a sufficiently wide inverse-cascading 
range, so that the enstrophy approaches its equilibrium value before 
finite-size effects interfere with the inverse cascade. It is also 
desirable to have a wide wavenumber range $k>s$, so that high 
Reynolds numbers can be achieved. For a given resolution, the choice of 
$K$ is a compromise between these two considerations. With the present 
choice of $K$, the enstrophy calculated for a $k^{-5/3}$ spectrum in 
the wavenumber interval $[3,50]$ is approximately 2\% less than that 
calculated for the same spectrum in the interval $[0,50]$. Hence, the 
energy transfer rate, which is calculated when the energy peak reaches 
wavenumber $k\approx3$, may be greater than its projected value by a 
few percents (assuming no significant finite-size interference to the 
inverse cascade up to that point in time). Dealiased 
pseudospectral method is used with $1024^2$ and $2048^2$ total modes. 
Several different Reynolds numbers, corresponding to several different 
values of $\nu_1$, were simulated. The lower resolution was used for 
$\nu_1=7\times10^{-4};5\times10^{-4}$, and the higher resolution 
was used for $\nu_1=3.5\times10^{-4};2.5\times10^{-4};
1.25\times10^{-4};6.25\times10^{-5}$. For all simulations, the 
initial energy spectrum had a negligibly small peak at the forcing 
wavenumber ($\Psi_2(50)\approx10^{-10}$).

\begin{figure}[tbph]
\centerline{\includegraphics{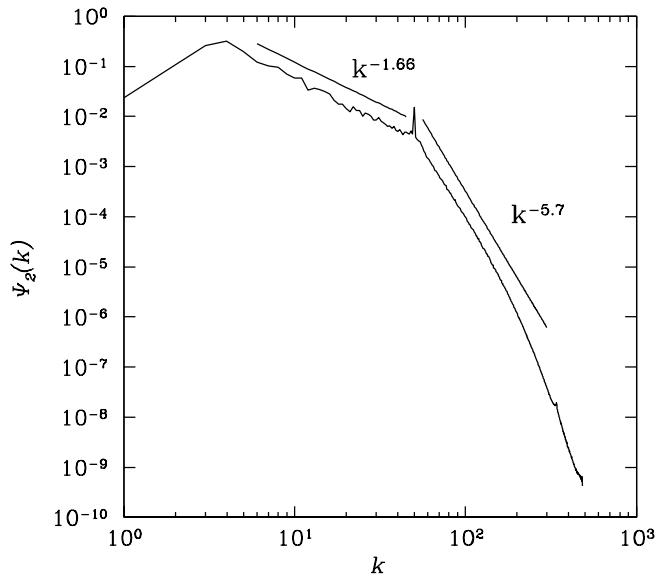}}
\caption{The energy spectrum $\Psi_2(k)$ vs. $k$ averaged between 
$t=10.2$ and $t=10.3$. The enstrophy averaged in the same period is
$\Psi_4=588$. The energy injection rate and viscosity coefficient 
used are $\epsilon=1$ and $\nu_1=7\times10^{-4}$, respectively. The 
inverse energy transfer rate is $\epsilon_0=\epsilon-2\nu_1\Psi_4=0.823$, 
which accounts for $82\%$ of the energy injection rate. The enstrophy 
range scales as $k^{-5.7}$. Hence, an inverse energy cascade carrying
$18\%$ of the energy injection rate is realized in the complete
absence of a direct cascade. The theoretical predictions are discussed 
in Section 4.}\label{CVT1phys1}
\end{figure}

\begin{figure}[tbph]
\centerline{\includegraphics{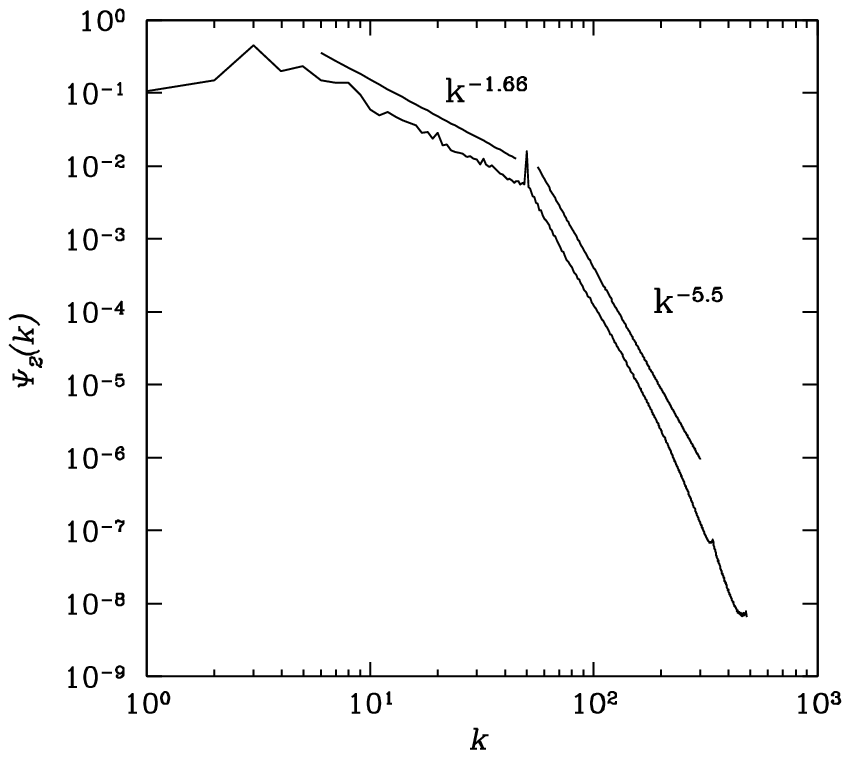}}
\caption{The energy spectrum $\Psi_2(k)$ vs. $k$ averaged between 
$t=8.8$ and $t=8.9$. The enstrophy averaged in the same period is
$\Psi_4=761$. The energy injection rate and viscosity coefficient 
used are $\epsilon=1$ and $\nu_1=5\times10^{-4}$, respectively. The 
inverse energy transfer rate is $\epsilon_0=\epsilon-2\nu_1\Psi_4=0.761$, 
which accounts for $76\%$ of the energy injection rate. The enstrophy 
range scales as $k^{-5.5}$. Hence, an inverse energy cascade carrying
$24\%$ of the energy injection rate is realized in the complete
absence of a direct cascade. The theoretical predictions are discussed 
in Section 4.}\label{CVT1phys2}
\end{figure}

For $\nu_1=7\times10^{-4}$, the energy peak was observed to reach 
$k\approx3$ at $t\approx10.2$, upon which both the energy spectrum 
(see Fig.~\ref{CVT1phys1}) and the total enstrophy were calculated 
(both time-averaged between $t=10.2$ and $t=10.3$). For a sense of 
this time scale, the dissipation time at the forcing wavenumber 
$s\approx50$ is $t_s=(2\nu_1s^2)^{-1}\approx0.3$. The calculated 
value of enstrophy $\Psi_4=588$ amounts to an energy dissipation rate 
$2\nu_1\Psi_4=0.823$, accounting for $82\%$ of the energy injection 
rate $\epsilon=1$. The energy growth rate $\epsilon_0$ is then $18\%$ 
of the energy injection rate, due to the inverse energy cascade.
It is evident from Fig.~\ref{CVT1phys1} that an inverse cascade 
is realizable in the complete absence of a direct cascade: the 
enstrophy range scales as $k^{-5.7}$, so that virtually all of the 
enstrophy is dissipated in the vicinity of the forcing region.
For $\nu_1=5\times10^{-4}$, the energy peak was observed to reach 
$k\approx3$ at $t\approx8.8$, upon which both the energy spectrum 
(see Fig.~\ref{CVT1phys2}) and the total enstrophy were calculated 
(both time-averaged between $t=8.8$ and $t=8.9$). The calculated value 
of enstrophy $\Psi_4=761$ amounts to an energy dissipation rate 
$2\nu_1\Psi_4=0.761$, accounting for $76\%$ of the energy injection 
rate $\epsilon=1$. The energy growth rate $\epsilon_0$ is then $24\%$ 
of the energy injection rate, due to the inverse energy cascade. 
The enstrophy range scales as $k^{-5.5}$, which is slightly shallower 
than its counterpart in the previous case. These results imply that
the inverse cascade and the enstrophy-range spectrum get stronger 
and shallower, respectively, as the viscosity coefficient $\nu_1$ 
decreases (the Reynolds number increases). Finally, for both cases,
the inverse cascade carries only a small fraction of the energy input 
to the largest scale and yet a $k^{-5/3}$ spectrum manifests itself
nonetheless. This suggests that a $k^{-5/3}$ inverse-cascading range
does not require the transfer of virtually all energy input to the
large scales.

\begin{figure}[tbph]
\centerline{\includegraphics{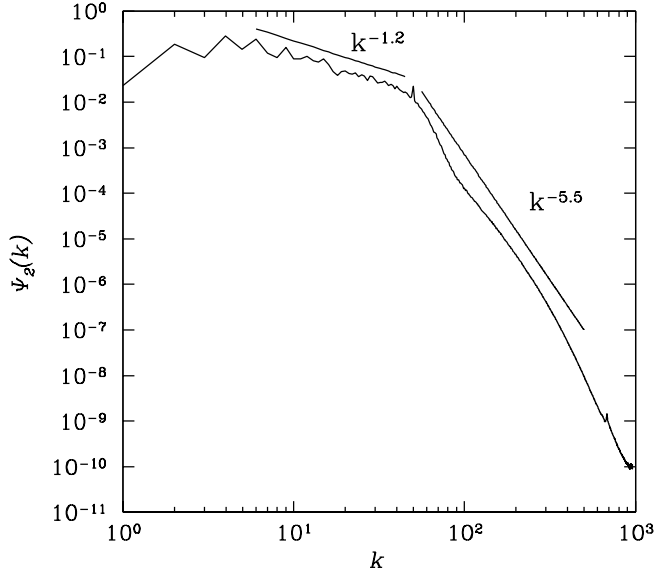}}
\caption{The energy spectrum $\Psi_2(k)$ vs. $k$ averaged between 
$t=7.74$ and $t=7.84$. The enstrophy injection rate and viscosity 
coefficient used are $s^2\epsilon\approx2500$ and $\nu_1=2.5\times10^{-4}$,
respectively. The energy range is significantly shallower 
than $k^{-5/3}$, due to a rise of the spectrum around $s$. This rise 
occurs (and persists) as the palinstrophy evolves within the 
insufficient-resolution range, after exceeding its equilibrium
value $\approx5\times10^6$ (achieving the value $5.26\times10^6$ 
at $t\approx1.2$) with an enstrophy-range slope $\approx-5$.}
\label{CVT1phys3}
\end{figure}

\begin{figure}[tbph]
\centerline{\includegraphics{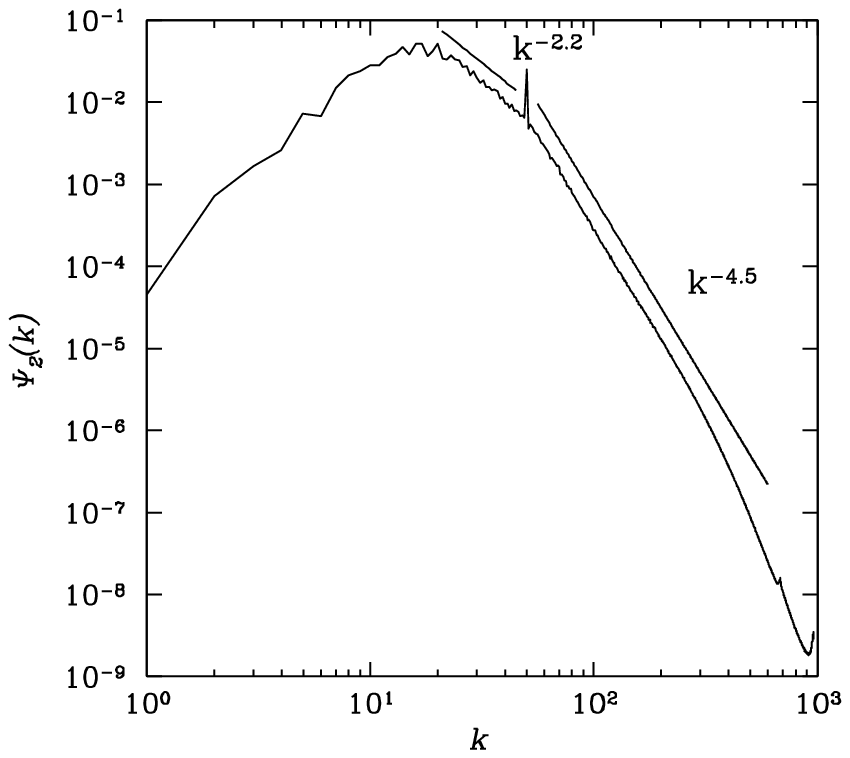}}
\caption{The energy spectrum $\Psi_2(k)$ vs. $k$ averaged between 
$t=1.1$ and $t=1.2$. The palinstrophy $\Psi_6$ averaged in the same 
period is $1.05\times10^7$. The respective enstrophy injection rate 
and viscosity coefficient used are $s^2\epsilon\approx2500$ and 
$\nu_1=1.25\times10^{-4}$, which amount to the palinstrophy equilibrium 
value $\approx1\times10^7$. One observes that the palinstrophy exceeds 
its equilibrium value well before a $k^{-5/3}$ spectrum is formed in 
the energy range. The enstrophy range is shallower than $k^{-5}$, 
clearly indicating insufficient resolution for the subsequent evolution.}
\label{CVT1phys4}
\end{figure}

The steep enstrophy-range spectra in Figs.~\ref{CVT1phys1} and 
\ref{CVT1phys2} seem to suggest that the present resolutions, 
particularly the higher one with $2048^2$ total modes, may be 
sufficient to resolve higher Reynolds numbers. It turns out, however, 
that further decrease of $\nu_1$ leads to spurious dynamics. For
example, for $\nu_1=3.5\times10^{-4};2.5\times10^{-4}$, the $k^{-5/3}$
inverse-cascading range is distorted after the energy peak has reached 
$k\approx8$ (at $t\approx3.5$): the energy range becomes shallower, 
due to a spectral rise at the forcing region. It is not known whether 
this phenomenon is due to finite-size or finite-resolution effects or 
both. (Such a rise is not observable in the previous simulations, 
where an inverse-cascading range is established down to $k\approx3$ 
without any significant departure from the expected $-5/3$ slope.) 
A reason to suggest that this phenomenon is mainly a manifestation 
of finite-resolution effects is that in these cases, the palinstrophy 
quickly grows to exceed its equilibrium value 
$\Psi_6=s^2\epsilon/2\nu_1$ with most of the (resolved) enstrophy 
range shallower than $k^{-5}$ (see Fig.~\ref{CVT1phys4}, 
for example), indicating insufficient resolution for the transient 
dynamics: a significant amount of palinstrophy would be carried 
by the unresolved (truncated) wavenumbers. The subsequent evolution 
of the enstrophy range, confined within the insufficient-resolution 
range, was observed to cause a rise of the spectrum around $s$. A 
growth of enstrophy is induced, thereby 
interfering with the inverse energy cascade. This phenomenon may be 
viewed as a redistribution of a significant palinstrophy amount back 
to the forcing region. This dynamical behaviour is illustrated by
Fig.~\ref{CVT1phys3}, from which the energy range is seen to
deviate considerably from the Kolmogorov--Kraichnan $k^{-5/3}$ spectrum. 

\begin{figure}[tbph]
\centerline{\includegraphics{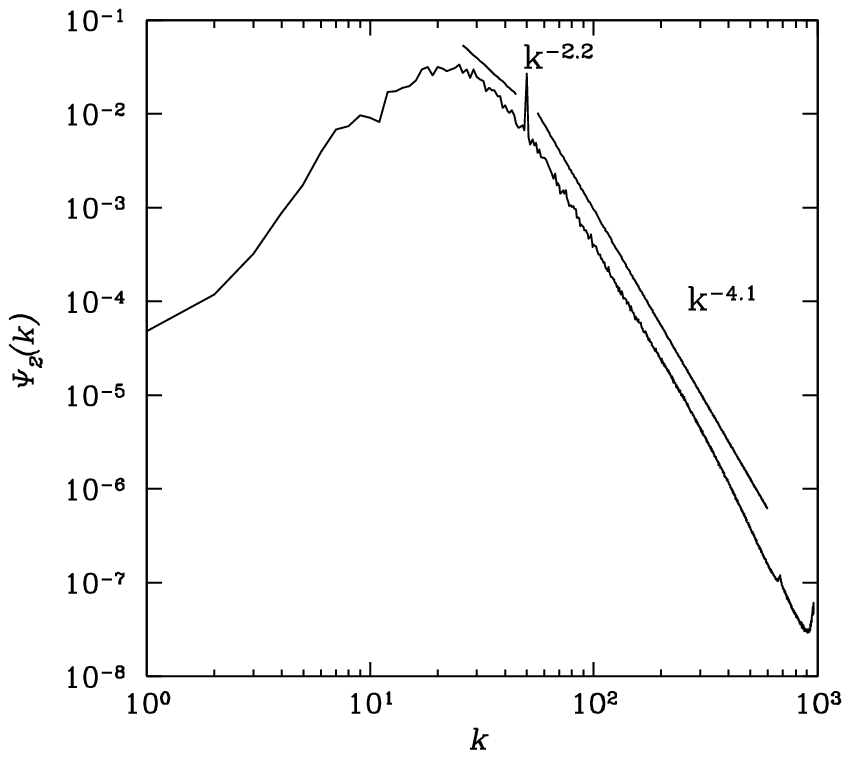}}
\caption{The energy spectrum $\Psi_2(k)$ vs. $k$ averaged between 
$t=0.90$ and $t=0.95$. The palinstrophy $\Psi_6$ averaged in the same 
period is $2.15\times10^7$. The respective enstrophy injection rate 
and viscosity coefficient used are $s^2\epsilon\approx2500$ and 
$\nu_1=6.25\times10^{-5}$, which amount to the palinstrophy equilibrium 
value $\approx2\times10^7$. One observes that the palinstrophy 
exceeds its equilibrium value well before a $k^{-5/3}$ spectrum is 
formed in the energy range. The enstrophy range is shallower than 
$k^{-5}$, clearly indicating insufficient resolution for the subsequent 
evolution.}\label{CVT1phys5}
\end{figure}

Higher Reynolds numbers seem to produce shallower enstrophy range
during the transient dynamics. For $\nu_1=1.25\times10^{-4}$ 
($\nu_1=6.25\times10^{-5}$), Fig.~\ref{CVT1phys4} 
(Fig.~\ref{CVT1phys5}) shows the energy spectrum averaged between 
$t=1.1$ ($t=0.90$) and $t=1.2$ ($t=0.95$), during which time about 
$0.7t_s$ ($0.3t_s$) has elapsed. 
The transient enstrophy range scales as $k^{-4.5}$ ($k^{-4.1}$) and 
extends to $k_\nu\approx500=10s$ ($k_\nu>10s$).\footnote{Since the
formation of an enstrophy range steeper than (but possibly arbitrarily 
close to) $k^{-3}$ requires virtually no enstrophy to be transferred to 
its higher-wavenumber end, the enstrophy range may acquire a slope 
between $-5$ and $-3$ during the transient stage. The width and slope 
of this range should depend on the Reynolds number and evolve with time.}
The enstrophy dynamics have become fully dissipative since the 
palinstrophy has already reached $\Psi_6(1.1)\approx1.05\times10^7$ 
($\Psi_6(0.9)\approx2.15\times10^7$), slightly exceeding its 
equilibrium value $s^2\epsilon/2\nu_1=1\times10^7$ 
($s^2\epsilon/2\nu_1=2\times10^7$). Clearly, the present resolution 
is insufficient, and the subsequent evolution is supposed to suffer 
from what have been termed ``finite-resolution effects'' as is seen 
in the case $\nu_1=2.5\times10^{-4}$. Apparently, the 
non-direct-cascading picture would not apply to this stage, nor would 
it apply to the subsequent dynamics if a persistent direct cascade 
could be established, withstanding any significant adjustments of 
the enstrophy-range spectrum. Rather, the realization of this picture 
requires that the enstrophy range relax toward a $k^{-5}$ spectrum. 
This relaxation would essentially involve a redistribution of 
palinstrophy (which has already achieved its equilibrium value 
$\Psi_6\approx s^2\epsilon/2\nu_1$) in the wavenumber range $k>s$. 
In some sense, the realizability of 
the non-direct-cascading picture requires that any direct-cascading 
states, regardless of the enstrophy-range slope, be ``unstable'' and 
that the palinstrophy be ``free'' to redistribute itself, until a 
$k^{-5}$ spectrum can be achieved. This dynamical description is 
supposed to be what subsequently occurs to the transient states 
depicted by Figs.~\ref{CVT1phys4} and \ref{CVT1phys5} (provided 
sufficient resolutions). In any case, the realization of transient 
direct-cascading dynamics, for $\nu_1=2.5\times10^{-4};6.25\times10^{-5}$,
suggests that for high Reynolds numbers, a transient direct-cascading 
stage would precede the non-direct-cascading picture if such a picture
could be realizable.

\section{Concluding remarks}

The nonlinear transfer and asymptotic dynamical behaviours of 
$\alpha$ turbulence have been investigated in this work. The results 
obtained include several estimates on the dynamical quantities, 
such as the dissipation of the inverse-cascading candidate, and 
various constraints on the spectral distribution of energy. The 
conservative transfer of quadratic invariants and the non-conservative 
transfer of other quadratic quantities are investigated. The possibility 
of a direct cascade in unbounded fluids is examined by exploring the 
constraint imposed by the conservative transfer of the advective term. 
A condition for the possible onset of direct cascade is derived, from 
which the dual-cascade hypothesis is reformulated. The realization of 
an inverse energy cascade in the complete absence of a direct enstrophy 
cascade in 2D NS turbulence is confirmed by numerical simulations.

For systems driven by a forcing spectrally localized at wavenumber 
$s$ (in the sense discussed in Section 3) and dissipated by 
a viscous operator $\nu_\mu(-\Delta)^\mu$, where $\nu_\mu>0$ and 
$\mu\ge0$, the invariants $\Psi_{2\alpha}$ and $\Psi_\alpha$ satisfy 
$\Psi_{2\alpha}\le s^\alpha\Psi_\alpha$. The nonlinear transfer 
destroys (creates) the quadratic quantity $\Psi_\theta$ for 
$\theta\in(\alpha,2\alpha)$ ($\theta\not\in[\alpha,2\alpha]$). This
phenomenon allows for vorticity to be created for systems with 
$\alpha<2$, which include the SQG equation. The enstrophy creation 
occurs in the high wavenumbers, and the created enstrophy is subject 
to viscous dissipation. For systems with $\alpha>2$, such as the RSF 
equation, energy is created. The creation of energy in RSF turbulence 
occurs in the low-wavenumber region. In the presence of a persistent 
inverse cascade, this energy creation would occur at an ever-growing 
rate. 

Fj{\o}rtoft-type arguments, based on the two conservation laws of 2D NS
turbulence, are reformulated for $\alpha$ turbulence, in an attempt to
gain more insights into the possibility of a direct cascade. It is 
shown that a weak inverse cascade, one that does not carry virtually
all of the injection of $\Psi_\alpha$ to ever-larger scales, cannot
co-exist with a direct cascade of $\Psi_{2\alpha}$. In order for a 
direct cascade to be permitted (not to say that it actually occurs), 
the inverse cascade is required to proceed toward the vanishing 
wavenumbers, carrying with it virtually all of the injection of 
$\Psi_\alpha$. This strong inverse cascade is widely believed to be 
what happens in the limit of infinite Reynolds number, according to 
the classical dual-cascade hypothesis. However, it is argued in this 
present paper that a dynamical picture that features no direct cascade 
for all Reynolds numbers is self-consistent and possible. This result 
considerably narrows down the possibility of a direct cascade and also 
explains the robustness of the inverse energy cascade observed in 
numerical simulations of 2D NS turbulence, regardless of what happens 
to the enstrophy \cite{Boffetta00,Borue94,Frisch84,Tran03c}. 
Furthermore, this hypothetical picture serves as a common dynamical 
background for both bounded and unbounded fluids. In both cases, no 
direct cascade would occur for all Reynolds numbers, and differences 
would arise only when the inverse cascade gets reflected by the largest 
available scale in the bounded case.

For $\alpha$ turbulence confined to a periodic domain, with a viscous 
dissipation operator $(-\Delta)^\mu$ ($\mu\ge0$) the only dissipation 
mechanism, no direct cascade is permitted. Moreover, an inverse cascade 
is not possible if the range of wavenumbers lower than the forcing 
wavenumber is sufficiently wide. More precisely, if the forcing 
wavenumber $s$ is no smaller than the geometric mean $\sqrt{k_0k_\nu}$ 
of the integral wavenumber $k_0$ (corresponding to the system size) 
and the dissipation wavenumber $k_\nu$ (the wavenumber beyond which 
the spectrum becomes steeper), then an inverse cascade is not possible. 
For $s<\sqrt{k_0k_\nu}$, an inverse cascade cannot be ruled out; however, 
such a cascade would only be marginal and characteristically different 
from the inverse cascade in the classical picture for unbounded systems. 
These results are the generalized versions of those obtained in 
\cite{Tran03a,Tran02a} for the NS system. They also apply to unbounded
fluids in equilibrium (if such an equilibrium could be achieved), where
$k_0$ is a wavenumber cutoff. 

Numerical analyses on 2D NS turbulence have been performed to confirm 
that a weak inverse energy cascade is realizable in the complete
absence of a direct enstrophy cascade. In particular, it is shown that
an inverse energy cascade, carrying approximately $24\%$ ($18\%$) of 
the energy injection to the large scales, is realized and accompanied 
by a $k^{-5.5}$ ($k^{-5.7}$) enstrophy range. The results also suggest
that the inverse-cascading strength (enstrophy-range spectrum) becomes 
stronger (shallower) as the Reynolds number increases and that for high 
Reynolds numbers, the non-direct-cascading picture (if it could be 
realizable) would be preceded by a transient direct-cascading stage. 
Unfortunately, due to resolution limitation, the numerics fall short 
on addressing quantitatively how the non-direct-cascading picture 
evolves with progressively higher Reynolds numbers. This open problem 
undoubtedly holds the key to a better understanding of the many aspects 
of finite-Reynolds number turbulence, possibly including the dual-cascade 
hypothesis in the limit of infinite Reynolds number. In the most favourable 
scenario, in which the KLB theory should turn out to be correct, a 
knowledge of this problem would be an asset for formulating a more 
complete theory of 2D NS turbulence. In particular, the new theory would 
have to address the dynamics corresponding to the (huge) spectral 
``gap'' between the critical $-5$ power-law scaling and the $-3$ 
power-law scaling of KLB theory (with or without a logarithmic correction). 
Intuitively, this ``gap'' would have to be filled with progressively 
higher Reynolds numbers, starting from the critical Reynolds number 
corresponding to the onset of a direct enstrophy cascade.

{\bf Acknowledgements}

The author would like to thank the two anonymous referees for their
constructive comments, which helped him clarify and improve the 
manuscript. He would also like to thank John Bowman for the numerical 
codes used in Section 7. This work was supported primarily by a Pacific 
Institute for the Mathematical Sciences postdoctoral fellowship. It 
was also supported in part by the Natural Sciences and Engineering 
Research Council of Canada through John Bowman's discovery grant.

%\bibliography{ref}

\end{document}